\documentclass[doublecol]{epl2}

\usepackage{amssymb}

\newcommand{\ri}{\mathrm{i}}
\newcommand{\dd}{\mathrm{d}}

\newcommand{\xx}{\mathbf{x}}

\newcommand{\sech}{\mathrm{sech}}

\title{Self-trapping of impurities in Bose-Einstein condensates:\\Strong attractive and repulsive coupling}

\shorttitle{Self-trapping of impurities in Bose-Einstein
condensates}

\author{M. Bruderer\inst{1} \and W. Bao\inst{2} \and D. Jaksch\inst{1}}
\shortauthor{M. Bruderer \etal}

\institute{
  \inst{1} Clarendon Laboratory - University of Oxford, Parks
Road, Oxford OX1 3PU, United Kingdom\\
  \inst{2} Department of Mathematics and Center for Computational Science and Engineering
  -- National\\
  University of Singapore, Singapore 117543}


\abstract{We study the interaction-induced localization -- the
so-called self-trapping -- of a neutral impurity atom immersed in a
homogeneous Bose-Einstein condensate (BEC).
Based on a Hartree description of the BEC we show that -- unlike
repulsive impurities -- attractive impurities have a singular ground
state in 3d and shrink to a point-like state in 2d as the coupling
approaches a critical value $\beta^\star$.
Moreover, we find that the density of the BEC increases markedly in
the vicinity of attractive impurities in 1d and 2d, which strongly
enhances inelastic collisions between atoms in the BEC.
These collisions result in a loss of BEC atoms and possibly of the
localized impurity itself.}

\begin{document}

\maketitle

Impurities immersed in liquid helium have proven to be a valuable
tool for probing the structure and dynamics of a Bose-condensed
fluid~\cite{Brewer-1966}. An example is the use of impurities for
the direct visualization of quantized vortices in superfluid
$^4$He~\cite{Bewley-NAT-2006}. Recently, the experimental
realization of impurities in a
BEC~\cite{Chikkatur-PRL-2000,Ciampini-PRA-2002} and the possibility
to produce quantum degenerate atomic
mixtures~\cite{Guenter-PRL-2006,Ospelkaus-PRL-2006} have generated
renewed interest in the physics of impurities. In particular, it was
pointed out that single atoms can get trapped in the localized
distortion of the BEC that is induced by the impurity-BEC
interaction~\cite{Lee-PRB-1992,Cucchietti-PRL-2006,Sacha-PRA-2006,Kalas-PRA-2006,Luhmann-AX-2007}.
More precisely, the impurity becomes self-trapped if its interaction
with the BEC is sufficiently strong to compensate for the high
kinetic energy of a localized state, an effect akin to the
self-trapping of impurities in liquid helium~\cite{Brewer-1966}.

However, there are two fundamental differences between liquid helium
and a BEC. First, for typical experimental parameters the healing
length of a BEC is three orders of magnitude larger than for liquid
helium~\cite{Dalfovo-JCP-2001}, and thus the so-called 'bubble'
model~\cite{Kuper-PR-1961} cannot be applied. A second important
difference is that the impurity-BEC interactions are tunable by an
external magnetic field in the vicinity of Feshbach
resonances~\cite{Ferlaino-PRA-2006,Klempt-PRA-2007}. In view of
recent experimental progress~\cite{Bloch-2007} it thus seems
possible to investigate the self-trapping problem in the same
physical system over a wide range of interaction strengths -- for
both attractive and repulsive impurities.

In the present paper we study the self-trapping properties of
impurities for strong attractive and repulsive impurity-BEC coupling
within the framework of a Hartree description of the BEC. This
necessitates the use of an essentially non-perturbative approach,
which is in contrast to previous analytical
studies~\cite{Lee-PRB-1992,Cucchietti-PRL-2006,Sacha-PRA-2006} where
the effect of the impurities on the BEC was treated as a small
perturbation. We first consider the effect of a highly localized
$\delta$-impurity on the BEC in one dimension. This approach
indicates that the density of the BEC is substantially increased in
the vicinity of an attractive impurity, which enhances inelastic
collisions and may result in the loss of the impurity atom. In
addition, we bring forward a scaling argument to show that
attractive impurity-BEC interactions can lead to a point-like ground
state of the impurity in 2d and 3d. Specifically, the ground state
is singular for arbitrarily small attractive coupling in 3d, which
seemingly contradicts the known perturbative results. Finally, in
order to extend our analytical findings we present numerical results
stemming from the exact solutions for the ground state of the
system, where unlike in previous numerical
studies~\cite{Kalas-PRA-2006,Luhmann-AX-2007} neither the impurity
nor the BEC are subject to an external trapping potential.

Our analysis of the self-trapping problem is based on the model
suggested by Gross~\cite{Brewer-1966,Gross-AP-1958} describing the
impurity wave-function $\chi(\xx,t)$ interacting with the condensate
wave-function $\psi(\xx,t)$ in the Hartree approximation. The model
is given by the coupled equations
\begin{eqnarray}
  \label{psi}
  \ri\hbar\partial_t\psi&=&-\frac{\hbar^2}{2m_b}\nabla^2\psi + \kappa|\chi|^2\psi +
  g|\psi|^2\psi\,,\\
  \label{chi}
  \ri\hbar\partial_t\chi&=&-\frac{\hbar^2}{2m_a}\nabla^2\chi +
  \kappa|\psi|^2\chi\,,
\end{eqnarray}
where the impurity-BEC interaction and the repulsive interaction
among the BEC atoms have been approximated by the contact potentials
$\kappa\delta(\xx - \xx^\prime)$ and $g\delta(\xx - \xx^\prime)$,
respectively. Here, $m_a$ is the mass of the impurity, $m_b$ is the
mass of a boson in the BEC and the coupling constants $\kappa$ and
$g>0$ depend on the respective s-wave scattering
lengths~\cite{Bolda-PRA-2003,Idziaszek-PRA-2006}. The functions
$\psi(\xx,t)$ and $\chi(\xx,t)$ are normalized as
\begin{equation}\label{norm}
    \int\dd\xx|\psi(\xx)|^2 = N\quad\mbox{and}\quad\int\dd\xx|\chi(\xx)|^2 = 1\,,
\end{equation}
where $N$ is number of atoms in the BEC. We tacitly assume that the
properties of the BEC, \emph{e.g.}~the density far away from the
impurity, are fixed. The interaction strength $\kappa$ on the other
hand is considered to be an adjustable parameter, since it may be
easily changed in experiments.

As a starting point we briefly review and systematically extend the
perturbative results on the self-trapping
problem~\cite{Lee-PRB-1992,Cucchietti-PRL-2006,Sacha-PRA-2006} based
on a variational approach. This method, however, yields results that
are of second order in $\kappa$, and hence is not adequate to
capture the differences between attractive and repulsive impurities.
In the main part of the paper, we present our non-perturbative and
numerical results in order to reveal these fundamental differences.
Finally, we conclude with a discussion of the physical implications
of our findings.

\section{Perturbative results}

To start we reformulate Eqs.~(\ref{psi}) and (\ref{chi}) in terms of
dimensionless quantities for a homogeneous BEC with density $n_0$.
The two relevant length scales in the model are the healing length
$\xi = \hbar/\sqrt{g n_0 m_b}$ and the average separation of the
condensate atoms $s = n_0^{-1/d}$, where $d$ is the dimension of the
system. Introducing the dimensionless quantities $\tilde{\xx} =
\xx/\xi$, $\tilde{\psi} =\psi\,s^{d/2}$ and $\tilde{\chi} =
\chi\,\xi^{d/2}$ one obtains the time-independent equations
\begin{eqnarray}
  \label{psix}
  \tilde{\psi}&=&-\frac{1}{2}\nabla^2\tilde{\psi}
  + \beta\gamma^d|\tilde{\chi}|^2\tilde{\psi} + |\tilde{\psi}|^2\tilde{\psi}\,,\\
  \label{chix}
  \varepsilon\tilde{\chi}&=&-\frac{\alpha}{2}\nabla^2\tilde{\chi} +
  \beta|\tilde{\psi}|^2\tilde{\chi}\,.
\end{eqnarray}
Here, $\alpha = m_b/m_a$ is the mass ratio, $\beta = \kappa/g$ is
the relative coupling strength, $\gamma = s/\xi$ and $\varepsilon$
is the energy of the impurity. The energy of the condensate
$E_\mathrm{bec}$, the interaction energy $E_\mathrm{int}$ and the
kinetic energy of the impurity $E_\mathrm{kin}$ are given by
\begin{eqnarray}\label{ebec}
    E_\mathrm{bec}&=&\gamma^{-d}\int\dd\xx\Big(\frac{1}{2}|\nabla\psi|^2 - |\psi|^2 + \frac{1}{2}|\psi|^4\Big)\,,\\\label{eint}
    E_\mathrm{int}&=&\beta\int\dd\xx|\chi|^2|\psi|^2\,,\\\label{ekin}
    E_\mathrm{kin}&=&\frac{\alpha}{2}\int\dd\xx|\nabla\chi|^2\,,
\end{eqnarray}
where all energies are measured in units of $gn_0$ and we dropped
the tilde for notational convenience, as in the remainder of the
paper. Equations~(\ref{psix}) and (\ref{chix}) have a trivial
solution with  $|\psi|^2=1$ and $\varepsilon = \beta$, which
corresponds to a delocalized impurity.

\subsection{Thomas-Fermi approximation}

If the density of the BEC changes smoothly we can apply the
Thomas-Fermi approximation, \emph{i.e.}~neglect the term
$\nabla^2\psi$ in Eq.~(\ref{psix}). In this regime we immediately
find that $|\psi(\xx)|^2 = 1 - \beta\gamma^d|\chi(\xx)|^2$, and thus
$\chi(\xx)$ obeys the self-focusing nonlinear Schr\"{o}dinger
equation
\begin{equation}\label{sfnlse}
  \varepsilon^\prime\chi = -\frac{1}{2}\nabla^2\chi -
  \zeta|\chi|^2\chi\,,
\end{equation}
with the self-trapping parameter $\zeta = \beta^2\gamma^d/\alpha$
and $\varepsilon^\prime = (\varepsilon-\beta)/\alpha$. In one
dimension Eq.~(\ref{sfnlse}) admits the exact
solution~\cite{Lee-PRB-1992,Juul-Rasmussen-PS-1986}
$\chi_\lambda(x) =(2\lambda)^{-1/2}\sech(x/\lambda)$, with the
localization length $\lambda = 2/\zeta$ and the energy
$\varepsilon^\prime = -\zeta^2/8$. In addition, numerical solutions
of Eq.~(\ref{sfnlse}) show that the wave-function of the form
$\chi_\lambda(\xx) = N_\lambda\sech(|\xx|/\lambda)$, with
$N_\lambda$ the normalization constant, is also an accurate
approximation to the exact solution in two and three
dimensions~\cite{Juul-Rasmussen-PS-1986,Anderson-PF-1979}. However,
these solutions can self-focus and become singular in finite
time~\cite{Juul-Rasmussen-PS-1986}.

A comparison of the individual terms in
$E_\mathrm{bec}+E_\mathrm{int}$ shows that the Thomas-Fermi
approximation is valid for $\zeta\ll 1$, which is a rather stringent
condition on the system parameters. However, the merit of
Eq.~(\ref{sfnlse}) lies in the fact that it yields an almost exact
wave-function for weakly localized impurities with
$\ell_\mathrm{loc}\gg 1$, where $\ell_\mathrm{loc}$ is the
localization length of the impurity. In addition, we see from the
solution of Eq.~(\ref{sfnlse}) that self-trapping occurs for
arbitrarily small coupling in 1d for both types of impurities.

\subsection{Weak-coupling approximation}

\begin{figure}[t]
\begin{center}
\includegraphics[width=6cm]{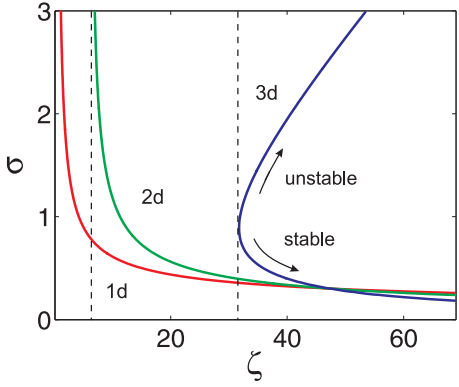}
\caption{The localization length $\sigma$ (solid lines) as a
function of the self-trapping parameter $\zeta$ obtained  by using a
Gaussian variational wave-function. Self-trapping occurs for
arbitrarily small $\zeta$ in 1d, for $\zeta>2\pi$ in 2d and for
$\zeta\gtrsim31.7$ in 3d (vertical dashed lines). From the two
self-trapping solutions for a given $\zeta$ in 3d, only the more
localized state is stable.} \label{figure.1}
\end{center}
\end{figure}

An alternative approach is to linearize the equation for $\psi(\xx)$
by considering small deformations $\delta\psi(\xx) = \psi(\xx) - 1$
of the
condensate~\cite{Lee-PRB-1992,Cucchietti-PRL-2006,Sacha-PRA-2006}.
This corresponds to a consistent expansion of Eqs.~(\ref{psix}) and
(\ref{chix}) in $\beta$, which results in the linear equations
\begin{eqnarray}\label{linpsi}
  \big(-\frac{1}{2}\nabla^2 + 2\big)\delta\psi&=&-\beta\gamma^d|\chi|^2\,,\\\label{linchi}
  \big(-\frac{\alpha}{2}\nabla^2 +
  2\beta\,\delta\psi\big)\chi&=&(\varepsilon-\beta)\chi\,,
\end{eqnarray}
where we assumed that $\delta\psi$ is real for simplicity. It can be
seen from Eq.~(\ref{linpsi}) that the linearization of
Eq.~(\ref{psix}) is valid in the regime
$|\beta|\gamma^d/\ell_\mathrm{loc}^d\ll 1$. The solution of
Eq.~(\ref{linpsi}) is given in terms of the Green's function
$G(\xx)$ satisfying the Helmholtz equation. One finds by inserting
the solution of Eq.~(\ref{linpsi}) into Eq.~(\ref{linchi}) that
$\chi$ obeys the nonlocal nonlinear Schr\"{o}dinger equation
\begin{equation}\label{nlse}
    \Big(-\frac{1}{2}\nabla^2 - 2\zeta\int\dd\xx^\prime
    G(\xx -
    \xx^\prime)|\chi(\xx^\prime)|^2\Big)\chi=\varepsilon^\prime\chi\,.
\end{equation}
Equation~(\ref{nlse}) minimizes the energy functional $F[\chi] =
E_\mathrm{kin} + E_\mathrm{def}$, with
\begin{equation}\label{efunct}
    E_\mathrm{def} = -\beta^2\gamma^d\int\dd\xx\,\dd\xx^\prime|\chi(\xx)|^2 G(\xx -
    \xx^\prime)|\chi(\xx^\prime)|^2\,,
\end{equation}
which has been discussed in the context of liquid helium by Lee and
Gunn~\cite{Lee-PRB-1992}. We see from Eqs.~(\ref{ekin}) and
(\ref{efunct}) that the self-trapping parameter $\zeta$ is
proportional to $E_\mathrm{def}/E_\mathrm{kin}$, \emph{i.e.}~the
ratio between the potential energy gained by deforming the BEC and
the kinetic energy of the impurity.

In order to estimate the critical parameters for which self-trapping
occurs we insert the harmonic oscillator ground state
\begin{equation}\label{chitrial}
    \chi_\sigma(\xx) =\big(\pi\sigma^2\big)^{-d/4}\prod_{j=1}^d
    \exp(-x_j^2/2\sigma^2)\,,
\end{equation}
with $\sigma$ the harmonic oscillator length, as a variational
wave-function into the functional~$F[\chi]$. The resulting energy
$f(\sigma)$ is given by
\begin{equation}\label{gfunct}
    f(\sigma) = \frac{\alpha d}{4\sigma^2} - \beta^2\gamma^d h(\sigma)\,,
\end{equation}
with the functions
\begin{eqnarray}\label{gg}
     h_\mathrm{1d}(\sigma)&=&\frac{1}{2}\exp(2\sigma^2)\mathrm{erfc}(\sqrt{2}\sigma)\,,\\
     h_\mathrm{2d}(\sigma)&=&-\frac{1}{2\pi}\exp(2\sigma^2)\mathrm{Ei}(-2\sigma^2)\,,\\
     h_\mathrm{3d}(\sigma)&=&\frac{1}{\pi}\left[\frac{1}{\sqrt{2\pi}\sigma}-
     \exp(2\sigma^2)\mathrm{erfc}(\sqrt{2}\sigma)\right]\,,
\end{eqnarray}
where $\mathrm{erfc}(x)$ is the complementary error function and
$\mathrm{Ei}(x)$ the exponential integral. The impurity localizes if
$f(\sigma)$ has a minimum for a finite value of $\sigma$, which
depends only on the self-trapping parameter $\zeta$.

The explicit expression for $f(\sigma)$, which appears to be a novel
result, allows us to determine the self-trapping threshold,
\emph{i.e.}~the critical coupling strength above which self-trapping
occurs, and the size of the self-trapped state even for highly
localized self-trapping solutions as long as
$|\beta|\gamma^d/\sigma^d\ll 1$. Figure~(\ref{figure.1}) shows the
localization length $\sigma$ as a function of the self-trapping
parameter $\zeta$ in 1d, 2d and 3d. Specifically, in 1d we find by
asymptotically expanding $f(\sigma)$ in the limit $\sigma\gg 1$ that
there exists a self-trapping solution for arbitrarily small $\zeta$
with $\sigma= \sqrt{2\pi}/\zeta$. The same expansion yields a
critical value $\zeta_\mathrm{crit} = 2\pi$ above which
self-trapping occurs in 2d. In the 3d case, numerical minimization
of $f(\sigma)$ shows that the critical value is
$\zeta_\mathrm{crit}\simeq31.7$, where the corresponding state is
highly localized with $\sigma\simeq0.87$. Interestingly, there exist
two self-trapping solutions for a given $\zeta$ in 3d, however only
the more localized state is stable.

\section{Beyond perturbation theory}

\begin{figure}[t]
\begin{center}
\includegraphics[width=3.5cm]{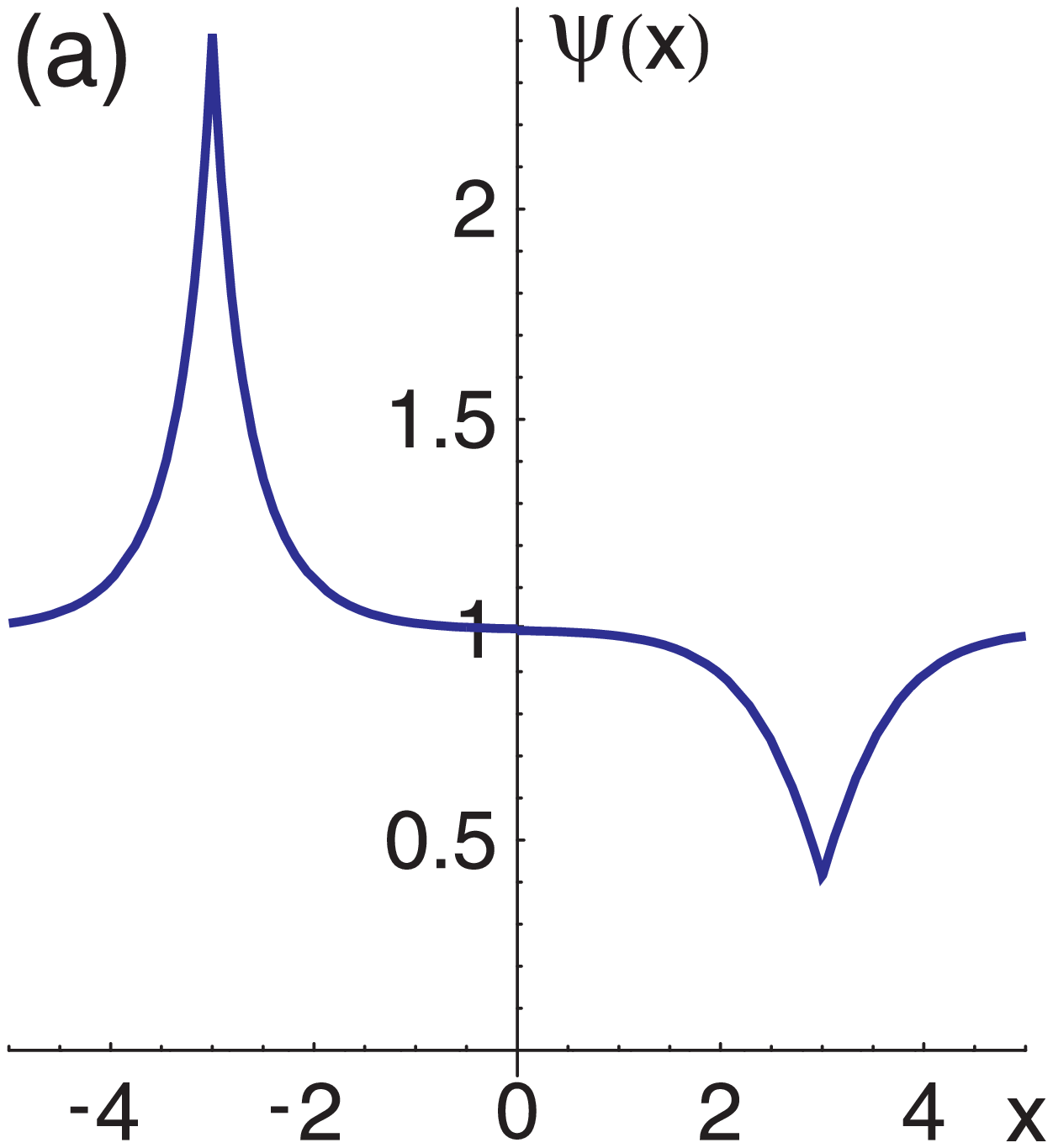}\hspace{0.35cm}\includegraphics[width=3.5cm]{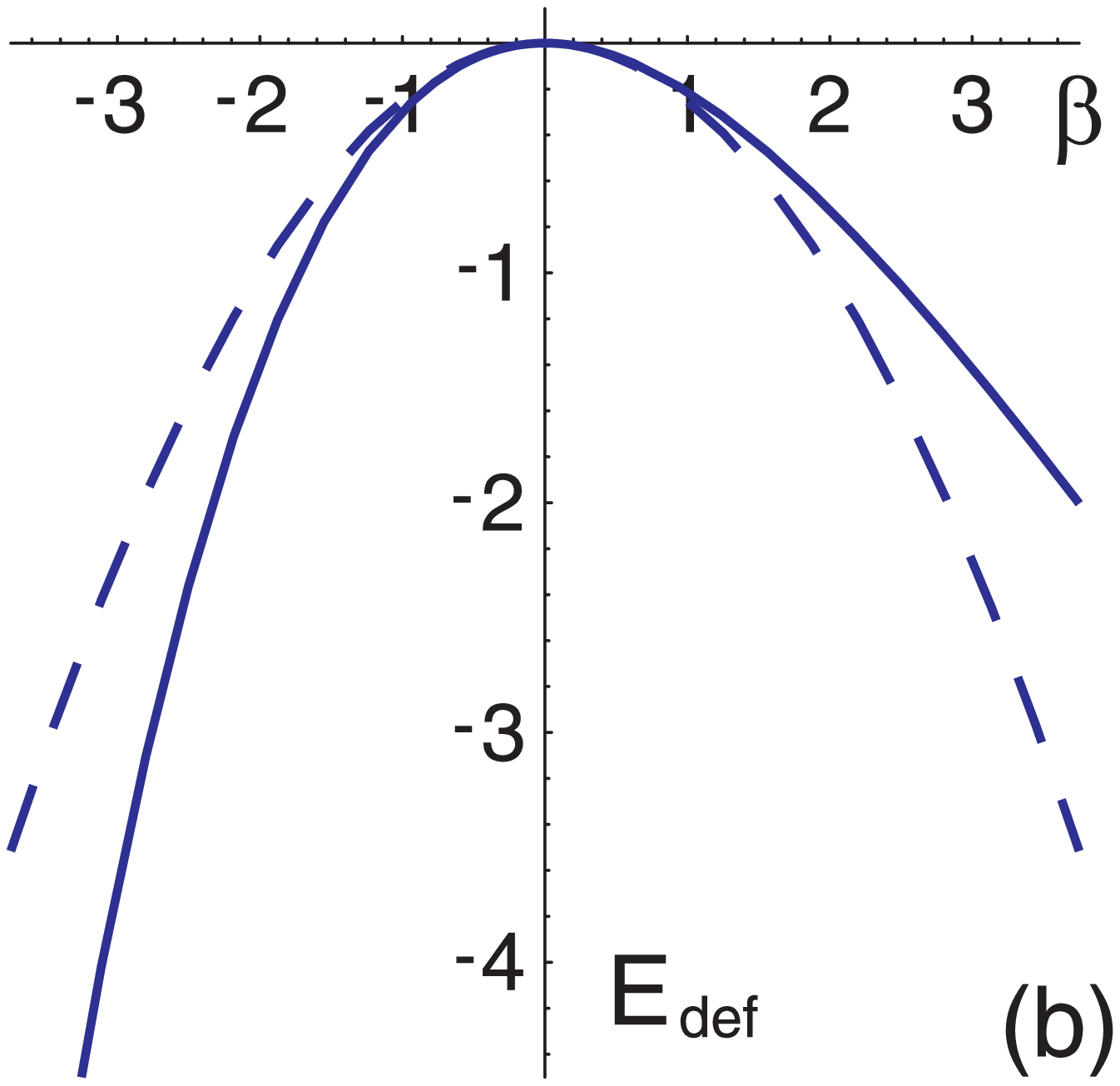}
\caption{(a) Deformation of the BEC for an attractive (left) and
repulsive (right) $\delta$-impurity with the same interaction
strength $\gamma|\beta|=1$.  (b) Deformation energy $E_\mathrm{def}$
for a $\delta$-impurity as a function of the interaction strength
$\beta$ (solid line) compared to the weak-coupling approximation
(dashed line) for $\gamma = 0.5$. Attractive (repulsive) impurities
have lower (higher) energy than predicted by the weak-coupling
approximation.} \label{figure.2}
\end{center}
\end{figure}

We now consider the self-trapping problem in the regime of strong
impurity-BEC interactions, \emph{i.e.}~the case of a localized
self-trapping state accompanied by a possibly strong deformation of
the BEC. We first illustrate the effect of higher order terms in
$\beta$ for a 1d system and subsequently discuss 2d and 3d systems
based on a scaling argument.

\subsection{The $\delta$-approximation in 1d}

In order to find approximate solutions of Eqs.~(\ref{psix}) and
(\ref{chix}) we assume that the impurity is highly localized due to
the strong impurity-BEC interaction. More precisely, we consider the
regime where the effect of the impurity on the BEC is essentially
the same as for a $\delta$-impurity with $|\chi(x)|^2=\delta(x)$.
This approximation is valid provided that $\ell_\mathrm{loc}\ll 1$,
\emph{i.e.}~the localization length is much smaller than the healing
length of the BEC, which is achieved for sufficiently large
$|\beta|$. Our approach focuses only on the deformation of the BEC
caused by the impurity. Nevertheless, the $\delta$-approximation
allows us to illustrate the differences between attractive and
repulsive impurities, which are also present in higher dimensions.

We first calculate the condensate wave-function $\psi(x)$ in
presence of a $\delta$-impurity at position $x_0$, which is
determined by the equation
\begin{equation}
  \label{psidelta}
  \left[-\frac{1}{2}\partial_{xx} - 1 + |\psi(x)|^2 + \beta\gamma\,\delta(x-x_0)\right]\psi(x) =
  0\,,
\end{equation}
with the boundary conditions $\psi(x) = 1$ and $\partial_x\psi(x) =
0$ in the limit $x\rightarrow\pm\infty$. Using the known solutions
of Eq.~(\ref{psidelta}) for $\beta=0$ and taking $x_0=0$ for
simplicity we obtain $\psi(x) = \coth(|x| + c)$ and $\psi(x) =
\tanh(|x| + c)$ for attractive and repulsive impurities,
respectively, with $c$ a constant. Given that the derivative of
$\psi(x)$ has a discontinuity at $x_0$ it follows that in both cases
$1 - [\psi(x_0)]^2 = \beta\gamma\,\psi(x_0)$ and thus
\begin{equation}\label{solz}
    \psi(x_0) = -\frac{\beta\gamma}{2} +
    \sqrt{1+\Big(\frac{\beta\gamma}{2}\Big)^2}\,,
\end{equation}
which fixes the constant $c$. We note that $\psi(x_0)\in(0,1]$ for
$\beta\geq 0$ and $\psi(x_0)\in(1,\infty)$ for $\beta < 0$. Thus the
density of the condensate $n(x_0)\equiv[\psi(x_0)]^2$ at the
position of the impurity $x_0$ is not bounded for attractive
interactions. The enhanced deformation of the BEC due to an
attractive $\delta$-impurity is already visible for a moderate
interaction strength as illustrated in Fig.~(\ref{figure.2})a.

The solution of Eq.~(\ref{psidelta}) allows us to determine the
deformation energy $E_\mathrm{def}$ to all orders of $\beta$. Using
Eqs.~(\ref{ebec}) and (\ref{eint}) we find that for both types of
impurities
\begin{eqnarray}\label{levelshift}
    E_\mathrm{def}=\frac{4}{3}\gamma^{-1}\bigg\{1-\bigg[1 +
    \Big(\frac{\beta\gamma}{2}\Big)^2\bigg]^{3/2}+
    \Big(\frac{\beta\gamma}{2}\Big)^3\bigg\}\,.
\end{eqnarray}
We note that the non-perturbative result for $E_\mathrm{def}$ is
consistent with the weak-coupling approximation since in the limit
$\sigma\rightarrow 0$ we find from Eq.~(\ref{gg}) that
$\beta^2\gamma h_\mathrm{1d}(\sigma)\rightarrow\beta^2\gamma/2$.
Figure~(\ref{figure.2})b shows the deformation energy
$E_\mathrm{def}$ as a function of the interaction strength $\beta$
in comparison with the weak-coupling result. It can be seen that
attractive impurities have lower energy than predicted by the
weak-coupling approximation, whereas repulsive impurities have
higher energy.

\subsection{Impurity collapse}

The results in the previous section indicate that the ground state
of the impurity and the BEC strongly depend on the sign of $\beta$.
We now show that the sign of $\beta$ has a crucial effect on the
ground state in 2d and 3d by using a scaling argument based on a
variational wave-function for $\psi(\xx)$ and $\chi(\xx)$.

Let us consider the total energy of the system $E_\mathrm{tot} =
E_\mathrm{bec} + E_\mathrm{int} + E_\mathrm{kin}$, which is clearly
bounded from below for $\beta>0$ but not necessarily for $\beta<0$.
We now analyze the scaling of the individual terms in
$E_\mathrm{tot}$ depending on the localization length of the
impurity and the deformation of the BEC. To this end we insert the
Gaussian trial function $\chi_\sigma(\xx)$ for the impurity and
$\psi_\sigma(\xx) = 1 + \delta\psi_\sigma(\xx)$ for the condensate
into $E_\mathrm{tot}$, where
\begin{equation}\label{trial}
    \delta\psi_\sigma(\xx) = \frac{a}{\sigma^{\delta/2}}\prod_{j=1}^d
    \exp(-x_j^2/b\sigma^2)\,,
\end{equation}
with $a$, $b$ and $\delta$ positive constants. In what follows we
are particularly interested in the limit $\sigma\rightarrow 0$ and
only consider finite deformations of the BEC. Consequently, we add
the constraint $\delta\leq d$ to assure that
$\int\dd\xx|\delta\psi(\xx)|^2\sim\sigma^{d-\delta}$ is finite. We
note that $\chi_\sigma(\xx)$ and $\psi_\sigma(\xx)$ have the correct
asymptotic behavior required for them to be valid physical states of
the system, however, they may not be a good approximation for the
ground state. We obtain that the individual energy contributions
scale as $E_\mathrm{int} \sim \sigma^{-\delta}$,
$E_\mathrm{kin}\sim\sigma^{-2}$ and
\begin{equation}
  E_\mathrm{bec} \sim c_0\sigma^{d - \delta -2} +  \sum_{j=1}^4 c_j\,\sigma^{d-j\delta/2}\,,
\end{equation}
where the $c_j>0$ depend on the system parameters.

Comparing the terms above one finds that for a three-dimensional
system $E_\mathrm{tot}$ is not bounded from below in the limit
$\sigma\rightarrow 0$ for a fixed $\delta$, \emph{i.e.}~the energy
of the system becomes arbitrarily low if the impurity collapses to a
point. Explicitly, in the 3d case we have $E_\mathrm{bec} \sim
c_0\sigma^{1 - \delta} + c_1\sigma^{3 - \delta/2} +  \cdots +
c_4\sigma^{3 - 2\delta}$ and hence a comparison of the exponents
yields that $|E_\mathrm{int}|
> E_\mathrm{kin} + E_\mathrm{bec}$ for $2<\delta<3$ in the limit
$\sigma\rightarrow 0$. Moreover, in the same limit
$\int\dd\xx|\delta\psi_\sigma(\xx)|^2\rightarrow 0$, \emph{i.e.}~the
deformation of the BEC is vanishingly small despite the collapse of
the impurity. We note that the above argument holds for arbitrarily
small $\beta$, and thus the weak-coupling approximation in 3d is
valid for $\beta>0$ only since for $\beta<0$ there is always a state
of lower energy with $\sigma\rightarrow 0$.

\begin{figure}[t]
\begin{center}
\includegraphics[width=6cm]{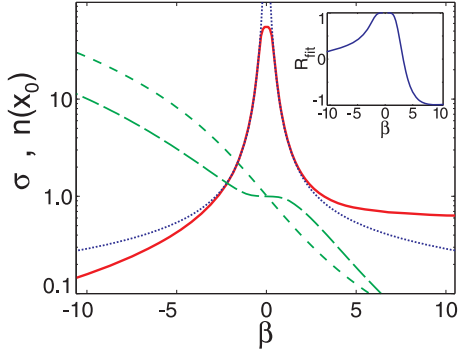}
\caption{The localization length $\sigma$ (solid line) and the
density of the condensate $n(x_0)$ (long-dashed line) at the
impurity position $x_0$ as a function of the interaction strength
$\beta$ in 1d. The weak-coupling approximation (dotted line)
accurately reproduces $\sigma$ in the regime
$|\beta|\gamma/\sigma\ll 1$. The localization length saturates to a
finite value $\sigma\sim 60$ as $\beta\rightarrow 0$ because of the
finite system size. The deformation of the BEC is qualitatively
described by $\delta$-approximation (short-dashed line). The inset
shows the smooth transition from a sech-type ($R_\mathrm{fit}=1$) to
a Gaussian ($R_\mathrm{fit}=-1$) self-trapping state for repulsive
impurities.} \label{figure.3}
\end{center}
\end{figure}

In the 2d case it is not possible to find a value for $\delta$ such
that the total energy $E_\mathrm{tot}$ is dominated by
$E_\mathrm{int}$ in the limit $\sigma\rightarrow 0$. Nevertheless,
for $\delta=2$ we obtain $E_\mathrm{int}\sim E_\mathrm{kin}\sim
E_\mathrm{bec}\sim\sigma^{-2}$, which implies that the collapse of
the impurity depends on the particular system parameters,
\emph{e.g.}~the coupling strength $\beta$. In contrast to the 3d
case this collapse is accompanied by a sizeable deformation of the
BEC since $\int\dd\xx|\delta\psi_\sigma(\xx)|^2$ is finite in the
limit $\sigma\rightarrow 0$ for $\delta=2$. Finally, for a
one-dimensional system similar considerations lead to result that
$|E_\mathrm{int}| < E_\mathrm{kin} + E_\mathrm{bec}$ in the limit
$\sigma\rightarrow 0$ for any valid $\delta$, and hence our
variational approach does not predict a collapse of the impurity.

\section{Numerical results}

\begin{figure}[t]
\begin{center}
\includegraphics[width=6cm]{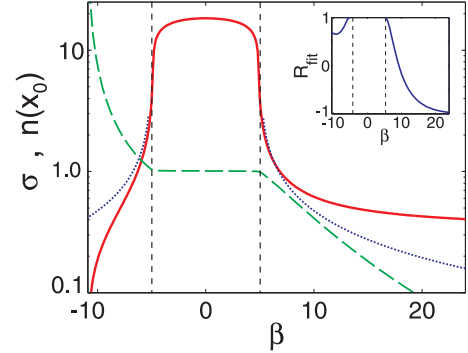}
\caption{The localization length $\sigma$ (solid line) and the
density of the condensate $n(x_0)$ (dashed line) at the impurity
position $x_0$ as a function of the interaction strength $\beta$ in
2d. The weak-coupling approximation (dotted lines) accurately
predicts the self-trapping threshold at $\beta_\mathrm{crit} =
\sqrt{8\pi}$ (vertical dashed lines). The localization length
saturates to a finite value $\sigma\sim 20$  for
$|\beta|<\beta_\mathrm{crit}$ because of the finite system size.
Attractive impurities shrink to a point-like state and the density
of the BEC $n(x_0)$ diverges as $\beta$ approaches a critical value
$\beta^\star\sim -10$ in agreement with our scaling argument. The
behavior of $R_\mathrm{fit}$ (inset) is similar to the 1d case.}
\label{figure.4}
\end{center}
\end{figure}

\begin{figure}[t]
\begin{center}
\includegraphics[width=5.2cm]{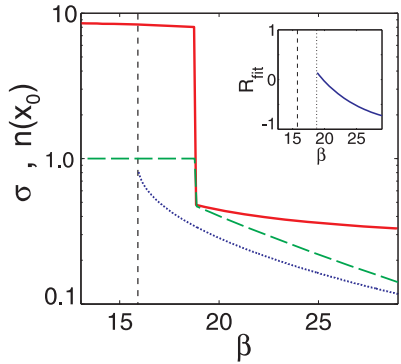}
\caption{The localization length $\sigma$ (solid line) and the
density of the condensate $n(x_0)$ (dashed line) at the impurity
position $x_0$ as a function of the interaction strength $\beta$ in
3d. There is no well-defined ground state for $\beta<0$ in 3d in
agreement with our scaling argument. The weak-coupling approximation
(dotted line) predicts a self-trapping threshold at
$\beta_\mathrm{crit} \simeq 15.9$ (vertical dashed line), which is
lower than the numerical result $\beta_\mathrm{crit} \simeq 18.8$.
The localization length saturates to a finite value $\sigma\sim 10$
for $\beta<\beta_\mathrm{crit}$ because of the finite system size.
The convergence of the impurity wave-function to a Gaussian
self-trapping state is slower than in 1d and 2d as indicated by
$R_\mathrm{fit}$ (inset).} \label{figure.5}
\end{center}
\end{figure}

We have determined the ground state of the impurity and the BEC
numerically by using the normalized gradient flow
method~\cite{Bao-SIAM-2004,Bao-MMS-2004} in order to extend the
analytical results. To make Eqs.~(\ref{psix}) and (\ref{chix})
amenable to computational modelling we assumed that both the
impurity and the condensate were enclosed in a sphere of zero
potential with infinitely high potential walls at its radius~$R$.
The system parameters for the numerical calculations were set to
$\alpha = 1.0$ and $\gamma = 0.5$, whereas $\beta$ was varied over a
large range.

A quantitative measure for the localization length
$\ell_\mathrm{loc}$ was found by fitting the functions
$\chi_\sigma(\xx)$ and $\chi_\lambda(\xx)$ to the exact impurity
wave function yielding the values $\sigma$ and $\lambda$. In
addition, we compared the corresponding fitting errors $S_\sigma$
and $S_\lambda$ and evaluated the quantity $R_\mathrm{fit} =
(S_\sigma - S_\lambda)/(S_\sigma + S_\lambda)$. Accordingly,
$R_\mathrm{fit} = -1$ indicates that the self-trapped state is
accurately described by a Gaussian $\chi_\sigma(\xx)$, whereas for
$R_\mathrm{fit} = 1$ the impurity is rather in a sech-type state
$\chi_\lambda(\xx)$. One would expect that $R_\mathrm{fit}\approx
-1$ for a highly localized impurity with $\ell_\mathrm{loc}\ll 1$
since the impurity sees only the harmonic part of the effective
potential $\beta|\psi(\xx)|^2$. On the other hand,
$R_\mathrm{fit}\approx -1$ for a weakly localized impurity with
$\ell_\mathrm{loc}\gg 1$ according to the Thomas-Fermi
approximation.

As can be seen in Fig.~(\ref{figure.3}), in the 1d case there is no
self-trapping threshold in agreement with the weak-coupling result,
which accurately reproduces the localization length $\sigma$ in the
regime $|\beta|\gamma/\sigma\ll 1$. However, we see that for large
$|\beta|$ attractive impurities are more localized than the
repulsive ones. Moreover, repulsive impurities undergo a smooth
transition from a sech-type to a Gaussian self-trapping state for
increasing $\beta$ as indicated by $R_\mathrm{fit}$. The deformation
of the condensate $n(x_0)$ at the position of the impurity $x_0$ is
qualitatively described by the $\delta$-approximation with a strong
increase in the density of the BEC for attractive impurities.

In two dimensions, the impurity localizes for
$|\beta|>\beta_\mathrm{crit}$, where $\beta_\mathrm{crit}$ is
correctly predicted by the weak-coupling result, as shown in
Fig.~(\ref{figure.4}). This is not accidental since the localization
length $\sigma$ diverges close to $\beta_\mathrm{crit}$ in 2d, as
can be seen in Fig.~(\ref{figure.1}), and thus the criterion
$|\beta|\gamma^2/\sigma^2\ll 1$ for the validity of the
linearization of Eqs.~(\ref{psix}) and (\ref{chix}) is always met
near $\beta_\mathrm{crit}$. In contrast, for large $|\beta|$ the
localization length $\sigma$ strongly deviates from the
weak-coupling result. In particular, attractive impurities shrink to
a point-like state and the density of the condensate $n(x_0)$
diverges as $\beta$ approaches a critical value $\beta^\star$, which
is in agreement with the scaling argument in the previous section.
The behavior of $R_\mathrm{fit}$ is similar to the 1d case.

For a 3d system we are able to obtain numerical results for
$\beta>0$ only, which in view of our scaling argument was to be
expected. The localization length $\sigma$ and the density of the
condensate $n(x_0)$ for repulsive impurities are shown in
Fig.~(\ref{figure.5}). We see that the weak-coupling approximation
underestimates the exact value $\beta_\mathrm{crit}$ since in 3d we
have that $\sigma\sim 1$ even close to $\beta_\mathrm{crit}$ and
thus $\beta\gamma^3/\sigma^3$ is not necessarily small. As shown in
the inset of Fig.~(\ref{figure.5}) the transition from a sech-type
to a Gaussian self-trapping state is slower than in 1d and 2d.

\section{Discussion and conclusion}

In our analytical and numerical investigation of the self-trapping
problem we have shown, in particular, that an attractive
impurity-BEC interaction leads to a strong deformation of the BEC in
1d and 2d and to a singular ground state of the impurity in 2d and
3d. However, both the high density of the BEC and the point-like
state of the impurity entail additional physical effects, which were
initially small and not taken into account in our model.

The high density of the BEC near the impurity enhances inelastic
collisions that lead to two- and three-body losses of the condensate
atoms~\cite{Roberts-PRL-2000}. Specifically, the inelastic
collisions might cause the loss of the impurity atom itself.
However, provided that the impurity is not affected by inelastic
collisions one can account for the loss of the condensate atoms by
adding  damping terms of the form $-\ri
\Gamma_2|\psi(\xx)|^2\psi(\xx)$ and $-\ri
\Gamma_3|\psi(\xx)|^4\psi(\xx)$ to Eq.~(\ref{psi}), where the
positive constants $\Gamma_2$ and $\Gamma_3$ are two- and three-body
loss rates, respectively. We note that these additional terms would
require the simulation of the full dynamics of the
system~\cite{Bao-SIAM-2003} and may have a nontrivial effect on the
self-trapping process.

Moreover, the contact potential approximation is valid as long as
the various length scales of the system, \emph{e.g.}~the
localization length $\ell_\mathrm{loc}$, are large compared to the
characteristic range of the exact interaction
potential~\cite{Bolda-PRA-2003}. Thus the interaction between the
condensate atoms and the impurity is no longer correctly described
by the contact potential $\kappa\delta(\xx - \xx^\prime)$ in the
limit $\ell_\mathrm{loc}\rightarrow 0$ where the influence of higher
partial waves becomes important. The resulting repulsion would of
course lead to a finite width of the impurity wave-function. We note
that with the above proviso our results are also valid for ionic
impurities with positive
charge~\cite{Cote-PRL-2002,Massignan-PRA-2005} in the regime where
the ion-boson scattering is dominated by the s-wave
contribution~\cite{Gribakin-PRA-1993,Cote-PRA-2000}.

Finally, we conclude with the remark that our model is also
applicable if the BEC is confined to a harmonic
trap~\cite{Kalas-PRA-2006} provided that the harmonic oscillator
length of the potential is much larger than the localization length
$\ell_\mathrm{loc}$ of the impurity and the healing length $\xi$ of
the BEC.

\acknowledgments

Part of this work was completed while M.B. was visiting the
Institute for Mathematical Sciences, National University of
Singapore, in 2007. This research was supported by the UK EPSRC
through QIP IRC (GR/S82176/01) and EuroQUAM project EP/E041612/1,
the EU through the STREP project OLAQUI, and the Singapore Ministry
of Education grant No. R-158-000-002-112.

\bibliographystyle{my_unsrt}
\bibliography{ThesisBib}

\end{document}